\documentclass[aps,prl,twocolumn,10pt,superscriptaddress,showpacs,showkeys]{revtex4-1}

\usepackage{amsmath,latexsym,amsfonts,amssymb}
\usepackage{natbib}
\usepackage{epsfig}
\usepackage{graphicx}
\usepackage{inputenc}
\usepackage{color}
\usepackage{setspace}
\usepackage{multirow}
\usepackage{array}
\usepackage{tabularx}
\usepackage{bm}
\usepackage{color}
\usepackage{xcolor}
\usepackage{gensymb}
\usepackage{xcolor}
\usepackage[left]{lineno}
\usepackage{float} 
\usepackage{ulem}
\usepackage{siunitx}

\newcommand{\bfs}{BaFe$_2$S$_3$\ }	
\newcommand{\bfse}{BaFe$_2$Se$_3$\ }	
\newcommand\st{\bgroup\markoverwith
  {\textcolor{blue}{\rule[.35ex]{5pt}{1.1pt}}}\ULon}
	
\begin{document}

%=============================================================================

\title{New insight on the phase diagram of the superconducting iron spin ladder BaFe$_2$S$_3$}

\author{Y. Oubaid}
\affiliation{Universit\'e Paris-Saclay, CNRS, Laboratoire de Physique des Solides, 91405, Orsay, France.}
\affiliation{Synchrotron SOLEIL, L\'\ Orme des Merisiers, Saint Aubin BP 48, 91192, Gif-sur-Yvette, France}

\author{V. Bal\'edent}
\affiliation{Universit\'e Paris-Saclay, CNRS, Laboratoire de Physique des Solides, 91405, Orsay, France.}
\email [Corresponding author: ] {victor.baledent@universite-paris-saclay.fr}

\author{O. Fabelo}
\affiliation{Institut Laue-Langevin, 38000 Grenoble, France}

\author{C. V. Colin}
\affiliation{Institut N\'eel, CNRS UPR~2940, 25 av. des Martyrs, 38042 Grenoble, France}

\author{S. Chattopadhyay}
\affiliation{UGC-DAE Consortium for Scientific Research, Mumbai Centre, 247C, 2nd Floor, Common Facility Building, BARC Campus, Trombay, Mumbai-400085, India}

\author{E. Elkaim}
\affiliation{Synchrotron SOLEIL, L\'\ Orme des Merisiers, Saint Aubin BP 48, 91192, Gif-sur-Yvette, France}

\author{A. Forget}
\affiliation{Universit\'e Paris-Saclay, CEA, CNRS, SPEC, 91191, Gif-sur-Yvette, France.}

\author{D. Colson}
\affiliation{Universit\'e Paris-Saclay, CEA, CNRS, SPEC, 91191, Gif-sur-Yvette, France.}

\author{D. Bounoua}
\affiliation{Universit\'e Paris-Saclay, LLB, CNRS, SPEC, 91191, Gif-sur-Yvette, France.}

\author{M. Verseils}
\affiliation{Synchrotron SOLEIL, L\'\ Orme des Merisiers, Saint Aubin BP 48, 91192, Gif-sur-Yvette, France}

\author{P. Fertey}
\affiliation{Synchrotron SOLEIL, L\'\ Orme des Merisiers, Saint Aubin BP 48, 91192, Gif-sur-Yvette, France}

\author{P. Foury-Leylekian}
\affiliation{Universit\'e Paris-Saclay, CNRS, Laboratoire de Physique des Solides, 91405, Orsay, France.}

\date{\today}

%=============================================================================
\begin{abstract}
\bfs and \bfse are the only two quasi-one-dimensional iron-based compounds that become superconductors under pressure. Interestingly, these two compounds exhibit different symmetries and properties. While more detailed and recent studies on \bfse using single crystals have advanced the filed towards a more universal description of this family, such a study is still lacking for the compound \bfs. Here, we present a detailed study of the crystalline and magnetic structure performed on single crystals using X-ray and neutron diffraction. We demonstrate a polar structure at room temperature within the $Cm2m$ space group, followed by a structural transition at 130 K to the polar $Pb2_1m$ space group. This space group remains unchanged across the magnetic transition at $T_N$=95 K, revealing multiferroic characteristics with a weak magnetoelastic coupling. The determined magnetic structure is monoclinic ($P_am$), with non-collinear magnetic moments, displaying a significant angle of 18° relative to the $a$-axis in the $(a,c)$ plane. This reexamination of the temperature-dependent properties of \bfs provides new insights into the physics of this system from multiple key perspectives.
\end{abstract}

\maketitle{}
%
% Intro

Unconventional superconductivity has been captivating physicists since its discovery in cuprates in 1986. From the copper age, we transitioned into the iron age during the 2000s with the discovery of a large number of families of iron based pnictides. A common feature of all these systems is their two-dimensional nature. More recently, it has been found that the dimensionality of these iron-based compounds can be further reduced to yield quasi-one-dimensional bands, forming ladder structures, and achieving superconductivity under pressure \cite{Takahashi2015, Yamauchi2015, Wu2019}. These compounds constitute a distinct family of iron chalcogenides with the general formula BaFe$_2$X$_3$ (where X is a chalcogen). In these systems, theoretical descriptions can be strongly facilitated by the low dimensional character of the electronic structure. The ultimate goal is to evidence a possible universal character in the series to elucidate the mechanism underlying high $Tc$ superconductivity. This requires revealing the relevant parameters that describe these compounds, namely the active degrees of freedom and the couplings that cannot be neglected. In these family of materials, it has already been demonstrated that these relevant parameters are the crystal structure, which determines the symmetries of the Hamiltonian, and the magnetic order, which couples to the structure through mechanisms related to frustrated exchange interactions \cite{Roll2023}. 
%Concerning chemical substitution, identifying the relevant effect is more challenging, as it influences both the electronic and crystal structures, such as in the substitution of Cs or K on the Ba site, or Co and Ni on the Fe site \cite{papier_sur_ces_dopages}. Furthermore, unlike the previously mentioned superconductors?cuprates and pnictides?neither superconductivity nor even metallization, a prerequisite for superconductivity, is achieved through these substitutions. Indeed, only the isovalent substitution of Se with S allows for the observation of superconductivity under pressure. 
A comprehensive study of the properties of \bfse has enabled an understanding of its structural \cite{Dong2014, Zheng2020, Weseloh2022}, dielectric \cite{Du2020}, and magnetic \cite{Mourigal2015, Zheng2020, Zheng2023, Roll2023} properties, as well as their couplings, evidencing a multiferroic character. In particular, the identification of a magnetic phase similar to the one of \bfs, near the superconducting dome under pressure, has brought us closer to this quest for universality within this family \cite{Zheng2022}. However, a similar study on \bfs is still lacking. Yet, this is a necessary step to comprehend these two compounds in a unified formalism, as they exhibit not only superconductivity but also interesting magnetoelectric and magnetoelastic properties.

\bfs crystallizes in the $Cmcm$ space group with unit cell parameters $a=8.787(1)$ \AA, $b=11.225(1)$ \AA, and $c=5.288(1)$ \AA , at 300 K \cite{Takahashi2015}. Each unit cell contains two Fe ladders extending along the $c$-axis, composed of edge-sharing FeSe$_4$ tetrahedra. The two ladders are equivalent by translation, unlike the Se-containing compound, which crystallizes in the non-standard $Pbnm$ space group (using the same unit cell setting) \cite{hong1972crystal, Zhang2018}. \bfs is a Mott semiconductor with a small band gap of around 40-60 meV. Below $T_N=120 \pm 25$ K, a magnetic transition is observed via neutron diffraction experiments primarily performed on powder (PND) \cite{Takahashi2015, Chi2016}. In contrast to \bfse, which stabilizes an original block-like antiferromagnetic (AFM) order \cite{Caron2011} at ambient pressure, the magnetic order of \bfs is stripe-like, with a propagation wave vector $\mathbf{k}=(\frac{1}{2}, \frac{1}{2}, 0)$, similar to that observed for KFe$_2$S$_3$ \cite{HAN2018} and CsFe$_2$S$_3$ \cite{Chi2016}. The ordered moments align along the $a$-axis and reach 1.1 $\pm$ 0.1 $\mu_B$ per Fe atom at low temperature, which is significantly below the expected magnetic moment for a high-spin Fe(II) atom (S=2) of 4 $\mu_B$. This reduction in moment could be attributed to a disordered spin contribution (spin glass part) or an Orbital Selective Mott Phase \cite{Herbrych2018}. The exact magnetic structure remains to be determined through single-crystal studies, which could provide information on potential spin canting, as observed in \bfs \cite{Saporov2011, Zheng2022b}.

In this letter, we present a comprehensive study of the crystalline and magnetic structures as a function of temperature, combining X-ray and neutron diffraction on single crystals. Our results obtained provide a wealth of new and crucial information on this compound with unprecedented detail. Firstly, \bfs is non-centrosymmetric at room temperature, and its symmetry is further reduced below a new transition at $T_S=130$ K, which results in a primitive unit cell comprising two inequivalent ladders. This new unit cell is preserved down to 17 K, showing only a weak coupling with the magnetic order below $T_N$. Nevertheless, this confirms the multiferroic character of \bfs. The magnetic structure obtained is monoclinic and deviates from a perfectly collinear antiferromagnetic structure as reported in the literature \cite{Takahashi2015}, with magnetic moments significantly canted with respect to the $a$-axis in the (a,c) plane.

% Details experimentaux
\paragraph{Experimental methods.} \bfs crystals were grown from the melt using the self-flux method \cite{Amigo2021}, starting from a mixture of BaS (99.9\%), Fe (99.9\%), and S (99.999\%) powders with a BaS:Fe ratio of 1:2.05:3. Two grams of the reagents were pelletized, placed in a carbon crucible, and then sealed in an evacuated quartz tube with a partial pressure of 300 mbar of Ar gas. The quartz tube was placed in a vertical tubular furnace, annealed at 1100$^\circ$C for 24 hours, and then cooled down to 750$^\circ$C at a rate of 6$^\circ$C/h before finally cooling to room temperature. This process resulted in a centimeter-sized pellet consisting of intimately co-aligned millimeter-sized rod-shaped crystals along the $c$ direction.

We conducted single-crystal and powder X-ray diffraction experiments at the SOLEIL synchrotron on the CRISTAL beamline. Selected single crystals, typically tens of microns in size, were measured using a wavelength of $\lambda$ = 0.6706 \AA\ at 17, 95, 144, and 300 K. Powder samples were measured on a two-circle diffractometer using a wavelength of $\lambda$ = 0.58182 \AA. Additional details are presented in the dedicated section in Supplemental material\cite{SI}. Neutron diffraction experiments were performed at 2 K on a millimeter-sized single crystal using the D9 hot neutron four-circle diffractometer at the Institut Laue-Langevin (ILL), Grenoble, France, with $\lambda = 0.836$ \AA. Powder neutron diffraction experiments as a function of temperature were conducted on the D1B spectrometer at ILL with an incident wavelength of $\lambda = 2.52$ \AA. Refinements were carried out using the Jana software \cite{PetricekVaclav2014} for crystallographic structures and the Fullprof program \cite{Rodriguez-Carvajal1993} for magnetic structures.

\begin{figure}[htbp]
\includegraphics [width=0.9\linewidth, angle=0]{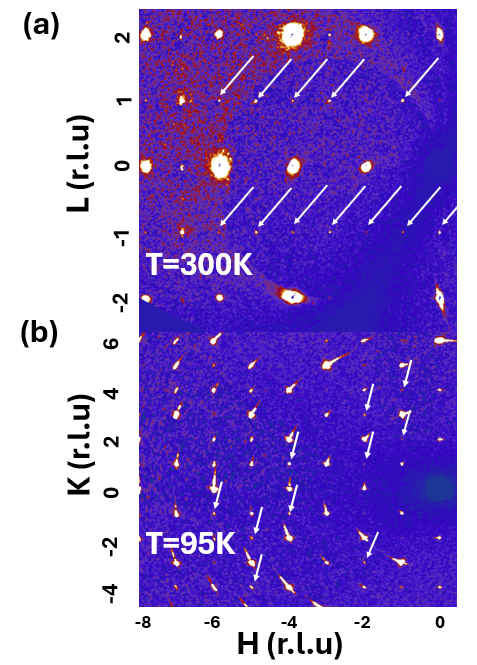}
\caption{\footnotesize(Color online) Reconstruction of the $(H0L)$ (a), $(H0K)$ reciprocal planes at 300 K and 95 K respectively. The arrows at 300K indicates the intensity at (H,0,L) with H+L odd, breaking the $c$ glide mirror perpendicular to the $b$-axis of the $Cmcm$ space group. The arrows at 300 K indicate the intensity at (H,K,0) with H+K odd, breaking the $C$ centering of the $Cm2m$ space group.}
\label{reconstruction}
\end{figure}

% Resultats RX
%300K et 140K
\paragraph{Results.} The 300 K x-ray pattern of \bfs already exhibits diffuse scattering segments along the $b$-axis, located on the Bragg lines, as visible in Fig.1 in the supplementary information \cite{SI}. Given their presence at all temperatures, these cannot be attributed to pre-transitional fluctuations, as might be expected for 1D materials. Instead, this is interpreted as stacking faults of the (a,c) planes (corresponding to the ladder plane) along the $b$ direction. As seen in Fig. \ref{reconstruction}a, we also observe intensity at $(h0l)$ positions with $l=2n+1$, typically three orders of magnitude smaller than other Bragg reflections, forbidden by the glide mirror {\it c} perpendicular to $b$. This cannot be attributed to multiple scattering since the extinctions associated with $C$ centering ($(hkl)$ with $h+k=2n$) remain intact, nor can it be caused by diffuse scattering from the faulted stacking along $b$. Similar observations were made at 140 K. Consequently, we conclude that the $Cmcm$ space group represents only an average structure. Among the three orthorhombic subgroups, only $Cm2m$ ($Amm2$ in the standard setting) is compatible with the extinction rules observed in the measurements. Notably, breaking the {\it c} glide mirror allows the two Fe-Fe bonds along the ladder to differ. The results of our refinement, presented in the supplemental information, show that this difference is insignificant within the uncertainties at 300 K (2.635(10) and 2.638(10) \AA) but a bit larger at 140 K (2.617(10) and 2.643(10) \AA \cite{Note}).

%95K et 17 K
At 95 K and 17 K, we observe a significant intensity, about 10\% of typical Bragg reflections, at positions forbidden by $C$ centering, specifically for $h+k=2n+1$, as visible in Fig. \ref{reconstruction}b. However, systematic extinctions remain observed at $(0kl)$ and $(0k0)$ positions for odd $k$, corresponding to the presence of a {\it b} glide mirror perpendicular to $a$ and a $2_1$ helical axis along $b$. Assuming this is a second-order structural transition, as we will demonstrate later, only $Pb2_1m$ ($Pmc2_1$ in the standard setting) is compatible with these symmetry elements among the four subgroups of $Cm2m$. The refined atomic positions in this space group for all temperatures are provided in the SI \cite{SI}. The comparison of the low temperature structures with the 140 K one shows that the two ladders of the unit cell rotate along their long axis but in opposite ways. This leads to the loss of the $C$ centering as for BaFe$_2$Se$_3$. The Fe displacements of about 0.02 \AA \ are accompanied by Ba movements of 0.04 \AA.  

\begin{figure}[htbp]
\includegraphics [width=0.99\linewidth, angle=0]{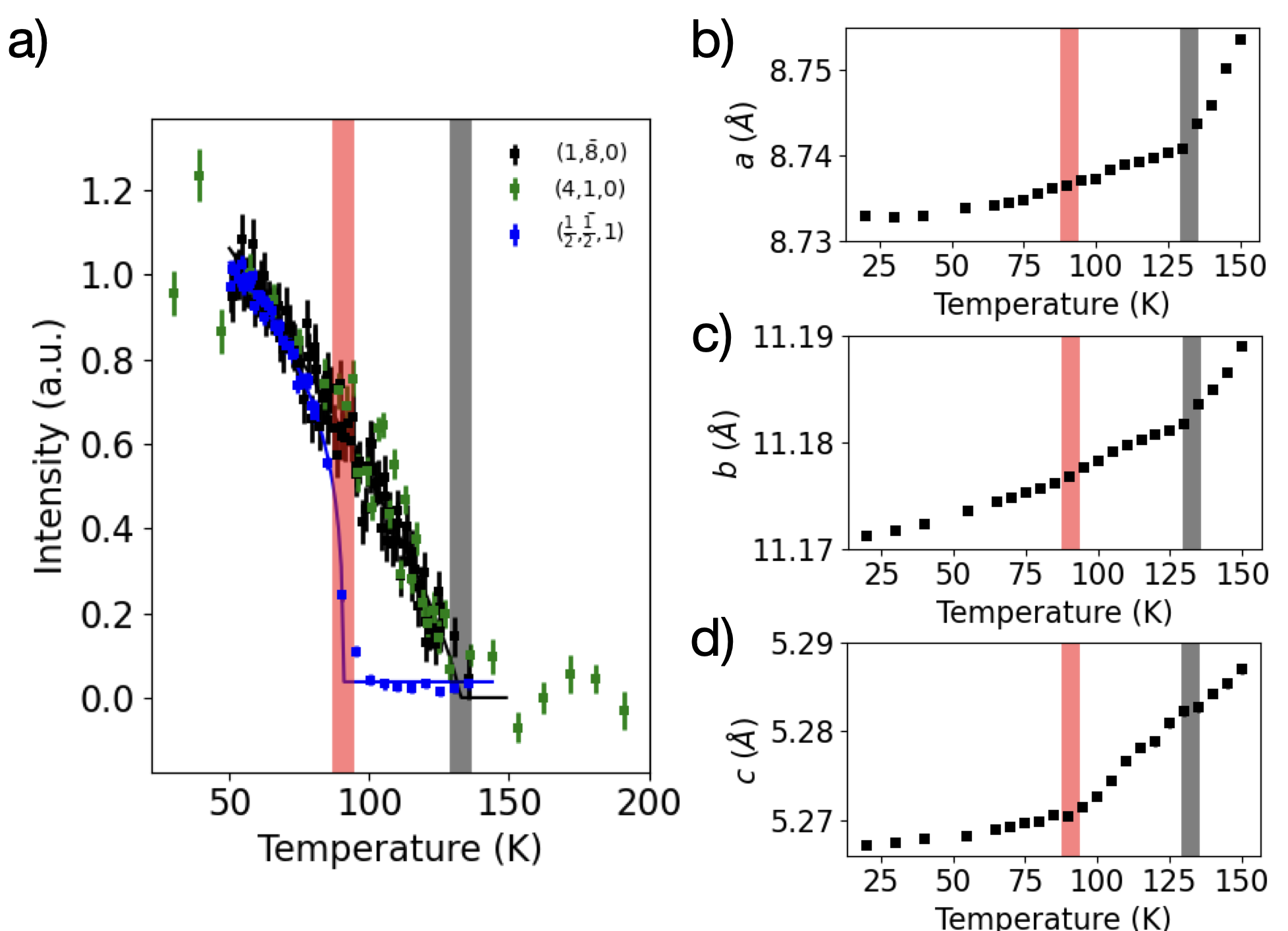}
\caption{\footnotesize(Color online) (a) Thermal variation of the intensities of a magnetic reflection ($\frac{1}{2},\bar{\frac{1}{2}},0$) and two nuclear reflections associated with the structural transition ($1,\bar{8},0$) (1 4 0) measured by neutron diffraction. Lines are fit with second order parameter functions. (b-d) temperature evolution of the lattice parameters a, b and c from X-ray powder diffraction using the position of the (2,0,0), (0,6,0) and (0,0,2) Bragg reflections respectively. Red and Black bands correspond to the fitted temperature for $T_N$ and $T_S$ in (a).}
\label{Fig3}
\end{figure}

%neutrons
The parent compound \bfse exhibits strong magneto-elastic coupling \cite{Zheng2020, Weseloh2022}, which can be an important parameter to understand the properties of these family of compounds. We have thus examined the case of \bfs and in particular the effect of the magnetic transition on the structural properties using neutron diffraction. To this end, we measured the evolution of a magnetic Bragg peak $(\frac{1}{2} \bar{\frac{1}{2}} 1)$ and two nuclear Bragg reflections breaking the $C$-centering characteristic of the structural transition from $Cm2m$ to $Pb2_1m$, $(1 \bar{8} 0)$ and  $(4 1 0)$ using single crystal neutron diffraction. As shown in Fig. \ref{Fig3}a, the temperature dependence of these peaks, proportional to the square of the magnetic and structural  order parameters (moment and atomic displacement), respectively, indicates a second-order transition. The obtained critical temperature and critical exponent values are $T_N$=94$\pm$3 K and $\beta$=0.16$\pm$0.1 for the magnetic transition, and $T_S$=130$\pm$5 K and $\beta$= 0.35$\pm$0.1 for the structural transition. It is noteworthy that a weak magnetic intensity persists above $T_N$, likely due to pretransitional fluctuations expected in a low-dimensional system. Furthermore, no visible anomaly at $T_N$ was observed in the evolution of the nuclear peak within the error bars, suggesting weak magneto-elastic coupling in this compound. This finding is corroborated by our temperature-dependent powder neutron diffraction measurements, where no anomalies in lattice parameters were detected accross the magnetic transition. Surprisingly, the lattice parameters also show no significant changes at the structural transition within the error bars.

To obtain better resolution in lattice parameters thermal evolution through both the structural and magnetic transitions, we performed powder diffraction experiments using X-rays. We report the evolution of the lattice parameters as a function of temperature, derived from the position of the $(200)$, $(060)$ and $(002)$ Bragg peaks, in Fig. \ref{Fig3}b,c,d. With this improved precision, we observe a change in slope around the structural transition temperature $T_S$=130 K, and an anomaly near the magnetic transition temperature $T_N$=95K mainly visible along the $c$ direction meaning the ladder axis. This confirms the expected behavior at the structural transition and indicates a weak but non-zero magneto-elastic coupling.

We then extended our study of \bfs by collecting 139 magnetic reflections from a single crystal to resolve the magnetic structure at 2 K. We performed a refinement using the average space group $Cmcm$ to reduce the number of parameters and constrain the fit. This approach is justified by the fact that the distortions between the $Cmcm$ and $Pb2_1m$ structures are sufficiently small not to affect the magnetic exchange interactions. A symmetry analysis starting from the $Cmcm$ group with a propagation vector ($\frac{1}{2},\frac{1}{2},0$) proposes four irreducible representations, of which only one provides a satisfactory refinement (see details in the supplementary information \cite{SI}). The resulting magnetic structure shows a zero component of the magnetic moment along the $b$-axis within experimental error bars. We fixed this component to zero in the final refinement to further constrain the fit. The absolute scaling was determined using 54 nuclear Bragg peaks measured in parallel, allowing the scale factor to be adjusted and yields an ordered moment of 1.33$\pm{0.1}$ $\mu_B$, which is comparable to previously published data from powder measurements \cite{Caron2011}.

\begin{figure}[htbp]
\includegraphics [width=1.0 \linewidth, angle=0]{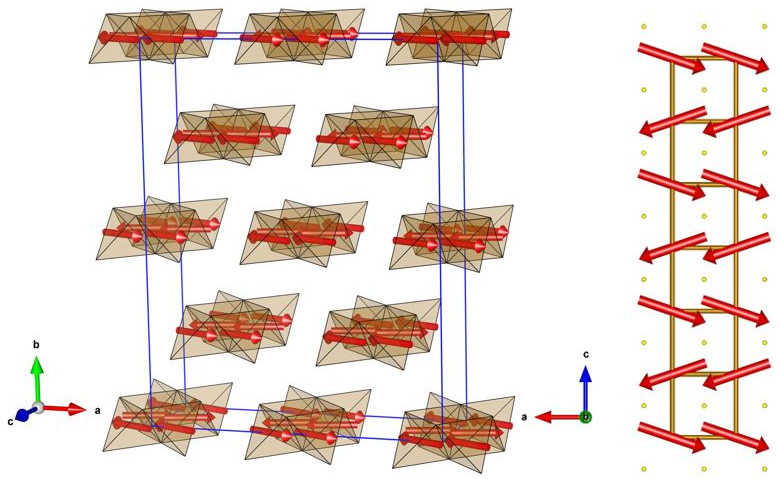}
\caption{\footnotesize(Color online) left : Magnetic structure at 2 K in the $P_am$ space group. The Fe ladders only are represented. right : projection along the $b$-axis, of the spins within the ladder}
\label{fig:mag_structure}
\end{figure}

The results of our refinement are summarized in Table \ref{tab:mag_refinement}, and the resulting magnetic structure is shown in Fig. \ref{fig:mag_structure}. The symmetries of the obtained magnetic structure are compatible with the monoclinic Shubnikov group $P2_1/m1'a$ (denoted as $P_a2_1/m$ in BNS notation). However, this group is not a subgroup of $Pb2_1m$, the real space group. The maximal magnetic group compatible with the parent structure is thus $P_am$ which leads to the presence of two inequivalent iron sites. We observed that the refinement of these two moments independently leads to values which are identical within the error bars. Consequently, we can confirm that from a strickly symmetry point of view, the magnetic space group should be $P_am$ but with a $P_a2_1/m$ quasi-symmetry. In contrast to the structure determined from powder samples \cite{Takahashi2015}, the magnetic moments are not perfectly aligned along the $a$-axis but are tilted by an angle of approximately 18\degree within the (a,c) plane. While the two moments of a rung remain collinear and ferromagnetically ordered,  two successive moment along the leg are antiferromagnetically ordered and not perfectly collinear. This leads to a slightly distorted stripe-like magnetic structure. Regarding the inter-ladder arrangement, we observe that the ladders are antiferromagnetically aligned along the a+b diagonal and ferromagnetically aligned along the a-b diagonal (see Fig. \ref{fig:mag_structure}), highlighting the monoclinic nature of this structure.

\begin{table}[htbp]
\caption{\footnotesize Positions of the Fe ions and their moments in \bfs at 2 K using the $\Gamma1$ irreps (see Supp. Mat. \cite{SI}) in the orthorhombic 8.74999 \AA,  11.22500 \AA, 5.28800 \AA unit cell ($R_{obs}$=14.6\%). The magnetic moment of $M_y$ has been fixed to zero. } 
\label{tab:mag_refinement}
\begin{tabular}{ccccccc}
\hline\hline
Atom & x & y & z & $M_x$ & $M_y$ & $M_z$\\ 
 \hline
$\rm Fe1\_1$ & 0.34569 & 0.5 & 0.0 & 1.26(3) & 0.0 & 0.42(4) \\ 
$\rm Fe1\_2$ & -0.34561 & -0.5 & 0.5 & -1.26(3) & 0.0 & 0.42(4) \\ 
$\rm Fe1\_3$ & -0.34569 & -0.5 & 0.0 & -1.26(3) & 0.0 & 0.42(4) \\ 
$\rm Fe1\_4$ & -0.34561 & 0.5 & 0.5 & 1.26(3) & 0.0 & 0.42(4) \\ 
\hline
\end{tabular}
\end{table}

%Firstly, one can observe that within the ladder structure, the iron atoms are connected in both directions (along \textbf{a} and \textbf{c}) by a Fe-S-Fe sulfur bridge, which is slightly shorter along the ladder direction (\textbf{c} axis). The angle of the Fe-S-Fe bridge along \textbf{c} is $70.4 ^\circ$, while it is $72.3 ^\circ$ along the a-axis.

%\begin{figure}[htbp]
%\includegraphics [width=1.0 \linewidth, angle=0]{Structure_Vs_T}
%\caption{Evolution of FeS$_4$ tetraedra angles and Fe-Fe dimerization along the ladder  as function of temperature.}
%\label{fig:FeS4_Vs_T}
%\end{figure}

\paragraph{Discussion.} 
The temperature-dependent properties of \bfs have been thoroughly revisited putting the physics of this system into perspective on several key aspects. 
Firstly, the discovery of a structural transition at $T_S$=130 K separated from the $T_N$ magnetic ordering, suggests reconsidering the origin of the anomaly in dielectric and transport properties at this temperature, which had previously been attributed to the magnetic transition \cite{Yamauchi2015, Chan2020}.
Secondly, the polar nature of the $Cm2m$ space group at room temperature persists below the structural transition temperature $T_S$ and within the magnetic phase below $T_N$=95 K. This endows the compound with multiferroic characteristics. With its relatively small band gap \cite{Yamauchi2015}, the electric polarization of \bfs expected from its non-centrosymmetric space group is difficult to measure, likely screened by conduction electrons \cite{Chan2020}. The nature of its multiferroicity, therefore, lies at the interface between classical magnetoelectric insulating compounds \cite{Gd_ME} and metallic multiferroic materials with novel fundamental physics and potentially unique properties \cite{Gui2022}. 
Thirdly, the determined magnetic structure not only does differ from the block-like magnetic order of \bfse but also strongly differs from the previously published classical antiferromagnetic structure published \cite{Caron2011}. Specifically, the magnetic moments are non collinear along the ladder, with a significant moment component along the ladder direction. The spin orientation within the (a,c) plane indicates that the local crystallographic anisotropy does not play the same significant role as in \bfse. Moreover, the strict ferromagnetic order between moments along the rung and between next nearest neighbour moments along the ladder is the indication of a strong modification of the exchange interactions compared to the not collinear umbrella-like spin structure of \bfse \cite{Zheng2022b}. 
The weak magneto-structural coupling of \bfs can be understood in light of what is known about the parent compound \bfse. In the latter, it has been demonstrated that the block magnetic structure originates from magnetic frustration caused by a next-nearest-neighbor coupling between Fe atoms within the same ladder \cite{Roll2023}. Not only does this strong frustration couple the structural and magnetic orders through the exchange-striction mechanism, but the resulting chiral magnetic structure adds an additional component to this coupling via the inverse Dzyaloshinskii-Moriya interaction \cite{Zheng2022b}. In the case of \bfs, the stripe-like antiferromagnetic structure suggests the presence of a weaker next-nearest-neighbor interaction, indicating less frustration. This explains the weak magneto-elastic coupling reported here.

 Additionally, the inequivalent inter-ladder coupling along the orthorhombic diagonals (a+b) and (a-b) results in a monoclinic magnetic structure with the space group $P_am$. However, no signature of a monoclinic crystal structure was observed in our diffraction data, which could be due to the weak magneto-elastic coupling.

\paragraph{Conclusion.} 
These results provide a new foundation for studying this compound and its related materials. Notably, investigating the lattice dynamics of this compound could provide insights not accessible via diffraction, similar to what has been done in \bfse \cite{Weseloh2022}. It also calls for a reexamination of the actual structure under pressure, particularly regarding the persistence of the non-centrosymmetric character into the superconducting phase, potentially paving the way for a one-dimensional non-centrosymmetric superconductor.

\paragraph{Aknowledgment.} We acknowledge SOLEIL for providing the synchrotron beamtime (proposals 		20230993, 20231465 and 20240327) and ILL for neutron diffraction (DOI: 10.5291/ILL-DATA.5-41-1215, DOI: 10.5291/ILL-DATA.5-31-2963, DOI: 10.5291/ILL-DATA.CRG-3079). This work was financially supported by the ANR COCOM 20-CE30-0029, the France 2030 programme ANR-11-IDEX-0003 via Integrative Institute of Materials from Paris-Saclay University - 2IM@UPSaclay, the Chinese Scholarship Council project (No. 201806830111) and the French Neutron Federation (2FDN).

\bibliography{Biblio.bib}

\begin{thebibliography}{27}
\bibitem [{Not()}]{Note}%
  \BibitemOpen
  \href@noop {} {\ }\bibinfo {note} {Although the incertainities from a single
  numerical refinement seem small, the real experimental errors are much larger
  ($\pm$ 0.01 \AA). They have been estimated statistically from a large set of
  different refinements of the same data}\BibitemShut {NoStop}%
\end{thebibliography}

\end{document}